\documentclass[aps,prd,twocolumn,superscriptaddress]{revtex4-1}

\usepackage{graphicx}
\usepackage{color}
\usepackage{amsfonts, amsmath}
\usepackage{caption}
\newif\ifoneauthor
\oneauthortrue

\begin{document}

\title{Folding Gravitational-Wave Interferometers}


\author{J. R. Sanders}
\email[]{jrsander@syr.edu}
\affiliation{Department of Physics, Syracuse University, NY 13244, USA}
\author{Stefan W. Ballmer}
\email[]{sballmer@syr.edu}
\affiliation{Department of Physics, Syracuse University, NY 13244, USA}



\date{\today}

\begin{abstract}
The sensitivity of kilometer-scale terrestrial gravitational wave interferometers is limited by mirror coating thermal noise. We explore the effect of folding the arm cavities of such interferometers. While simple folding alone does not reduce the  mirror coating thermal noise, it makes the folding mirror the critical mirror, opening up a variety of design and upgrade options.
\end{abstract}

\pacs{42.79.Bh, 95.55.Ym, 04.80.Nn, 05.40.Ca}

\maketitle

\section{Introduction}

The recent detection of gravitational waves from binary black hole coalescences marks the beginning of an exciting new era in gravitational wave astronomy and astrophysics. The two signals observed by Advanced LIGO in its first observing run have proven the existence of binary stellar-mass black hole systems and set strong constraints on deviations from general relativity in the strong-field regime \cite{GW150914,GW151226}. During the first observing run, Advanced LIGO was at less than half of its design range, sensitive to neutron star binary mergers at 75 Mpc.  Planned improvements in sensitivity will increase the range of Advanced LIGO to 200 Mpc, corresponding to an increase in observable volume and event rate by a factor of 18.

The generation of interferometers in operation and under construction are kilometer-scale dual-recycled Fabry-Perot Michelson interferometers, with noise performance primarily limited by quantum noise and coating Brownian thermal noise. Quantum noise reduction through squeezed vacuum states is an active area of development, and systems are being prepared for future integration into Advanced LIGO. Thermal noise is of concern in the most sensitive band of ground-based interferometers, between $50-350~\rm{Hz}$, and is a limiting noise source for upgrades to the current generation of gravitational-wave interferometers (Advanced LIGO \cite{Harry2010}, 
Advanced Virgo \cite{TheVirgo:2014hva} and Kagra \cite{Somiya:2011np}).
The importance of the effect for gravitational wave detectors has driven a theoretical \cite{Braginsky1999, Braginsky2000, Levin1998, PhysRevD.63.082003, Levin2008, Levin2009, Evans2008, PhysRevD.90.043013} and experimental \cite{Rao2003, Harry:2001iw, PhysRevLett.89.237402, Harry:06, Harry2010, Cole2013} interest in understanding and improving the fundamental thermal noise of optical elements. Schemes for extending interferometer range beyond Advanced LIGO are under active investigation, including methods for noise reduction in the current interferometer facilities \cite{Miller2015} and the impact of lengthening the interferometer arms in larger-scale successor facilities \cite{PhysRevD.91.082001,ETDesign}.

In this work, we propose a novel topology for interferometer sensitivity improvement and thermal noise mitigation that can be integrated into existing interferometer facilities. Introducing a folding mirror into the Fabry-Perot arms of an Advanced LIGO-like interferometer doubles the effective arm length of the optical cavity. The potential sensitivity advantage is canceled by the additional coating thermal noise
contribution of the folding mirror, which becomes the critical mirror for thermal noise. Thermal noise mitigation techniques already under development, such as the monocrystalline film coatings and cryogenic test masses discussed below, can be applied to the folding mirror at a smaller scale than that required for thermal noise reduction in an unfolded optical cavity. Reducing the thermal noise contribution from this folding mirror alone reduces the impact of thermal noise on the whole interferometer significantly, resulting in a net range increase.

\section{Review of Thermal Noise Reduction Techniques}

Thermal noise in interferometer test masses is tightly linked to energy dissipation through the fluctuation-dissipation theorem \cite{Callen1951,PhysRevD.42.2437,Levin1998,Levin2008}. The dominant source of thermal noise in gravitational wave interferometers is mechanical loss in the high-index layers of the dielectric optical coatings, known as coating Brownian thermal noise. The direct connection to mechanical loss is evident in the expression for the thermal noise power spectral density \cite{Levin1998}, given by
\begin{equation}
\label{eq:thermalPSD}
S_{\rm{Br,coat}} = \frac{4k_B T}{\pi f} U \phi(f) .
\end{equation}
Here $U$ is the strain energy associated with a static pressure profile in the shape of the optical beam intensity on the mirror surface, normalized by the total driving force. $\phi(f)$ is the mechanical loss angle \cite{PhysRevD.42.2437}. 

Significant effort has gone into optimizing dielectric coatings for gravitational-wave detectors \cite{Harry:07}, resulting in silica-tantala ($\rm SiO_{2}-Ta_{2}O_{5}$) dielectric stacks with titania-doping ($\rm TiO_{2}$) in the $\rm Ta_{2}O_{5}$ layers. These coatings have excellent optical quality, with low optical losses and sub-nanometer roughness. Although the mechanical loss angle of the high-index $\rm Ta_{2}O_{5}$ layers is only about $\phi = 2.5 \times 10^{-4}$, this dominates the Brownian thermal noise of the mirror, and is the most limiting constraint in the design of new gravitational-wave detectors \cite{PhysRevD.91.082001}.

A number of ideas to improve the coating Brownian noise have been explored. Since the performance of a gravitational wave interferometer is measured in strain, rather than displacement, long arm cavities are the first line of defense against coating Brownian noise. First articulated by Weiss \cite{weiss:1972}, this is the primary reason for the kilometer-scale arm cavities of Advanced LIGO, and provides the most straightforward path for future gravitational wave detectors \cite{PhysRevD.91.082001}. However, extending the arm cavities is not an attractive option for upgrading current facilities.

Perhaps the most obvious approach to reducing the coating Brownian thermal noise is cryogenic operation of the interferometer test masses. This is currently actively pursued by the Japanese KAGRA project \cite{Somiya:2011np}. While promising the potentially biggest reduction in Brownian thermal noise, this approach has its challenges. The amplitude of the thermal noise scales with the square root of temperature, requiring relatively low mirror temperatures for significant improvements. This in turn poses a challenge for extracting the heat deposited by the interferometer beam. Other material properties, in particular the loss angle and Young's modulus, also have a temperature dependence, and can potentially negate any noise improvements. In particular the substrate material of choice for room temperature mirrors - fused silica ($\rm SiO_{2}$) - is no longer a good choice due to excess mechanical loss at cryogenic temperatures  \cite{cryochapter2012}. Finally, integrating the required low vibration heat pumps and cryo-shields into an interferometer requires a significant amount of engineering.

In principle, a simpler approach is to find new coating materials with reduced mechanical loss. There has been some progress in this direction, and in particular two types of coating seem promising: 
crystalline coatings based on substrate transferred GaAs/$\rm{Al}_x\rm{Ga}_{1-x}\rm{As}$ multilayers (AlGaAs) \cite{Cole2013}, and amorphous silicon coatings \cite{Murray2015}, both of which are further discussed below.

Finally, coating Brownian noise can be reduced by increasing the beam spots on mirrors, effectively averaging the fluctuations over a larger area. Increasing the arm cavity g-factor is a straightforward way to increase spot size. Advanced LIGO uses this strategy; the arm cavity g-factor of 0.8305 increases the beam spot area by more than a factor of two over its minimum size. Further increases in g-factor will be increasingly difficult, as alignment and optical quality requirements increase as the g-factor goes to 1. Using a different mode type, such as Laguerre-Gaussian beams \cite{PhysRevLett.105.231102,PhysRevD.92.102002}, increases the spot size for a given cavity g-factor. However, this introduces a degeneracy problem. Different mode shapes are resonant at the same time, making it difficult to achieve good contrast at the interferometer's dark port \cite{PhysRevD.84.102001}. 

\begin{figure}[htb]
  \centering
  \includegraphics[width=1 \columnwidth]{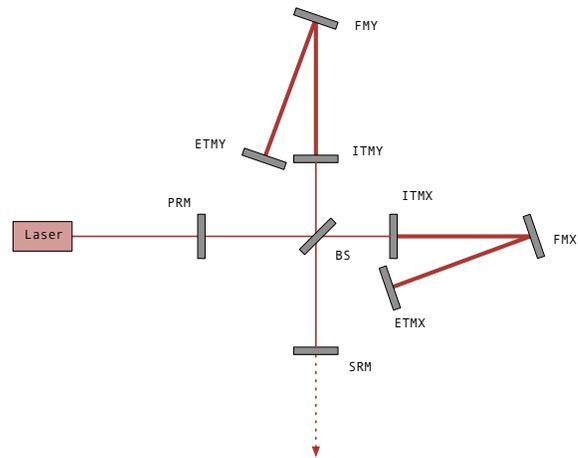}
  \caption{Core optics of proposed singly-folded Fabry-Perot interferometer. Note that the folding mirror is probed twice by the laser light.}.
\label{fig:foldedSketch}
\end{figure}

Folding the Fabry-Perot arm cavities is an alternative strategy to effectively sample many beam spots on the mirrors \cite{PhysRevD.88.062004}. This sampling is effective in the case of traveling-wave cavities, at the cost of requiring traveling wave recycling cavities. In the case of standing-wave cavities, Heinert et. al. \cite{Heinert2014} pointed out that a folding mirror incurs extra thermal noise compared to a straight reflection due to the interference fringe pattern. For coating Brownian noise, the excess in the power spectral density is $50\%$:
\begin{equation}
\label{eq:HeinertModifiedPSD}
S_{\rm{Br,coat,folded}} = \frac{3}{2}S_{\rm{Br,coat}}, 
\end{equation} 
\noindent where the $S_{\rm{Br,coat}}$ is the term assumed in equation \ref{eq:thermalPSD}. Any potential benefit from folding a cavity also has to overcome this excess noise.

To examine the impact of folding mirror thermal noise, consider a singly-folded Fabry-Perot cavity, as shown in Figure \ref{fig:foldedSketch}. Assume identical thermal noise properties and spot sizes for the input coupler (ITM), end mirror (ETM), and folding mirror (FM). The ITM and ETM noise power adds incoherently, while the FM contributes twice coherently to the round trip thermal noise. Compared to a non-folded Fabry-Perot cavity, this leads to a displacement thermal noise power increase of  
\begin{equation}
\label{eq:TNsimpleexample}
\frac{(1)_{\rm ITM} + (1)_{\rm ETM} +\left(2^2 {\frac{3}{2}}\right)_{\rm FM}}{(1)_{\rm ITM} + (1)_{\rm ETM}} = 4 . 
\end{equation} 
The corresponding increase of a factor two in amplitude thermal noise, together with a doubled cavity length, results in an identical strain noise of the folded cavity. In practice, the ITM has a thinner coating and slightly lower thermal noise than the ETM, and this simple folding results in a slight thermal noise increase over Advanced LIGO.

So why bother with folding a cavity in the first place? The key lies in the relative importance of the folding mirror thermal noise, as is evident in Equation \ref{eq:TNsimpleexample}.
Given the dominant contribution of folding mirror thermal noise to the total thermal noise in a singly-folded Fabry-Perot cavity, direct improvements to the thermal noise properties of the folding mirror can improve the total strain amplitude thermal noise by up to a factor of two. This might be particularly interesting for a cryogenic upgrade; the required cryo-baffles are easier to integrate around the isolated folding mirror, completely avoiding the complicated optical layout around the central interferometer. Introducing a single folding mirror to an interferometer that is otherwise similar to current gravitational wave detectors is an attractive intermediate option for improving detector sensitivity while introducing minimal novel technology.




 
 
 
 
 
\section{Options for Improving Folding Mirror Thermal Noise}




\subsection{Monocrystalline film coatings}

Compound semiconductor epitaxial multilayers, including aluminum gallium arsenide (AlGaAs), are candidates for high-Q coatings with high optical quality. AlGaAs-based microresonators have been shown to have quality factors up to 40,000 at room temperature \cite{Cole2012} , corresponding to a loss angle a factor of ten lower than for amorphous silica-tantala coatings \cite{Cole2013}. This loss angle could be decreased by a further factor of two for an operating temperature near the zero crossing in thermal expansion of single-crystal Si ($\sim$124 K) and up to an additional factor of 10 for temperatures of 10 K or below, with loss angles below $5 \times 10^{-6}$ measured in AlGaAs at liquid He temperatures \cite{Cole2012}.  


AlGaAs coatings have achieved excellent optical properties, with absorption $<$1 ppm measured at both 1064 and 1550 nm, and scatter loss at or below 3 ppm in the same range \cite{Cole2016}. However, the index of refraction of the layers making up the AlGaAs-based coating is higher than that of silica/tantala: GaAs has index $n = 3.48$ and $\rm{Al}_{0.92}\rm{Ga}_{0.08}\rm{As}$ has index $n = 2.94$. 

The fabrication process imposes the primary limitation in the adoption of AlGaAs coatings. AlGaAs-based crystalline coatings are initially grown on a flat GaAs substrate and must be transferred to the fused silica substrates used for aLIGO. The technology required for the adoption of crystalline coatings has been progressing rapidly, and coatings up to 20 cm in diameter could be immediately available for future upgrades. The current maximum coating diameter is limited by the size of commercially-available GaAs wafers. However, with additional investment, it is also technically feasible to generate transferred coatings with diameters approaching 40 cm.\cite{Cole2016email}. 

In the low-diffraction limit, the diffraction loss $L$ of a beam with radius $w$ incident on a mirror of radius $a$ is given by

\begin{equation}
\label{eq:LowDiffractionWaist}
L= \exp\left(-\frac{2w^2}{a^2}\right).
\end{equation}

\noindent For an AlGaAs coating of radius 0.1 m and an acceptable diffraction loss of $10\sim\rm{ppm}$ this gives a maximum beam size on the folding mirror of $w = 4.2\;\rm{cm}$. For comparison, the current aLIGO test masses are 34 cm in diameter, with a beam size on the end test mass of 6.2 cm \cite{aLIGO}. Using AlGaAs only on the folding mirror allows for larger beam sizes at the test masses while satisfying this constraint on the AlGaAs coating.

\subsection{Cryogenics}


Given that the loss angle of amorphous coatings remains relatively constant with temperature \cite{Yamamoto2006}, coating thermal noise should decrease proportionally with temperature. However, the implementation of cryogenic operations requires significant changes to the test masses due to the behavior of fused silica under cooling. The loss angle of the fused silica substrate used in aLIGO peaks sharply at 50K, mandating a change in test mass substrate for cryogenic operation \cite{cryochapter2012}. Although the sapphire substrates used for CLIO and KAGRA exhibit desirable mechanical loss changes with temperature, significant fabrication and integration challenges limit their utility for third-generation interferometers. Pure silicon is an appealing alternative; high quality silicon substrates can be produced using techniques developed for the semiconductor industry, and silicon ribbon suspensions can be produced to provide suspension thermal noise performance comparable to the silica fibers used in aLIGO \cite{Cumming2014}. As silicon is opaque to the 1064 nm light used in aLIGO, a change in laser wavelength to 1550 nm or longer would be required for transmitting optics. If silicon is used only for a folding mirror, the current
wavelength of 1064 nm would be acceptable.

Amorphous silica/tantala coatings may still be effective at cryogenic temperatures, as high quality coatings can be made for 1550 nm wavelengths with very low optical absorption. Ion-beam sputtered amorphous silicon ($\alpha$Si) has lower intrinsic mechanical loss and a higher index of refraction, allowing the same reflectivity to be attained with fewer dielectric layers than silica/tantala coatings \cite{Murray2015}. However, current $\alpha$Si coatings exhibit absorption at the level of 1000 ppm and require further development to improve optical quality. 

\subsection{Potential noise improvements}




The singly-folded Fabry-Perot interferometer can be implemented with relatively few changes to core interferometer systems. Maintaining these systems sets constraints on potential thermal noise mitigation, particularly for cryogenic methods. As the silicon substrates proposed for cryogenic operation are not transparent at the current 1064 nm operating wavelength, silicon substrates could only be used for the folding mirror. For similar reasons, amorphous silicon coatings will only be of interest for interferometers operating at 1550 nm. 

Consider three potential folding mirror thermal noise mitigation cases: cryogenic Si substrate with silica-tantala coating, room-temperature fused silica substrate with AlGaAs coating, and cryogenic Si substrate with AlGaAs coating. In all cases, the full expression for coating thermal noise \cite{Harry:07} was used to calculate the coating thermal noise contribution for each optic, with the coating thickness for the folding mirror calculated assuming a folding mirror with similar transmissivity to the end test mass ($t = 5 \times 10^{-6}$). For the cryogenic cases, the substrate temperature was taken to be 123 K \cite{InstrumentWhitePaper}. The input and end test masses are assumed to be room-temperature fused silica with silica-tantala coatings.

All cases exhibit reduced coating thermal noise over Advanced LIGO. Of particular interest is the results for AlGaAs coatings. Assuming a 20 cm diameter coating, a room temperature folding mirror has a 25\% reduction in coating thermal noise, while a cryogenic folding mirror reduces coating noise by nearly a factor of 2. If the test mass coatings were also replaced with AlGaAs coatings, a room-temperature interferometer would have coating thermal noise nearly a factor of 2 lower, and an interferometer with a cryogenic folding mirror would reduce thermal noise by a factor of nearly 4. The results of the calculation are summarized in Tables \ref{table:ASDresults} and \ref{table:moreASDresults}, in terms of the noise amplitude spectral density $S^{1/2}_{\rm{Br,coat}}$ at 100 Hz.

\begingroup
\begin{table}
\begin{tabular}{l|l|l}
\hline
\hline
Folding mirror &  $S^{1/2}_{\rm{Br,coat}}(100\;\rm{Hz})$ & \% of aLIGO \\
 \hline
 aLIGO baseline & $2.5 \times 10^{-24}$ & 100 \\
 \hline
 Silica-tantala (290 K) & $2.78 \times 10^{-24}$ & 111 \\
 \hline
 AlGaAs (290 K) & $1.94 \times 10^{-24}$ & 77.3 \\
 \hline
 Silica-tantala (123 K) & $2.05 \times 10^{-24}$ & 81.7 \\
 \hline
 AlGaAs (123 K) & $1.31 \times 10^{-24}$ & 52.1 \\
 \hline
\end{tabular}
\caption{Thermal noise amplitude spectral density at 100 Hz for varying folding mirror thermal noise mitigation schemes. In all cases, the thermal noise mitigation was applied only to the folding mirror. For each case, the operational wavelength was maintained at 1064 nm and the input and end test masses were assumed to be identical to aLIGO. The amplitude spectral density is reported for consistency with the sensitivity curves presented in Figure \ref{fig:SensitivityProjection}. \label{table:ASDresults}}
\end{table}
\endgroup

\begingroup
\begin{table}
\begin{tabular}{l|l|l}
\hline
\hline
Folding mirror & $S^{1/2}_{\rm{Br,coat}}(100\;\rm{Hz})$& \% of aLIGO \\
\hline
AlGaAs (290 K) & $1.62 \times 10^{-24}$ & 64.3 \\
\hline
AlGaAs (123 K) & $7.43 \times 10^{-25}$ & 29.6 \\
\hline
\end{tabular}

\caption{Thermal noise amplitude spectral density at 100 Hz for a folded interferometer with AlGaAs coatings on all three mirrors. For each case, the operational wavelength was maintained at 1064 nm. \label{table:moreASDresults}}
\end{table}
\endgroup

\section{Operational Considerations}
The success of the current generation of gravitational-wave detectors is both an opportunity and a complication for any future upgrade. A successful upgrade will increase the range and sensitivity of the detector, allowing detection of binary black hole signals at higher redshift and improving detection capability for novel sources. However, upgrade installation and commissioning will take the detectors offline, halting the collection of data for the duration. Introducing folded Fabry-Perot arms to current interferometer facilities is a relatively low-impact proposition. Assuming no change in laser frequency, the central interferometer, input, and readout require minimal modifications. The potential payoff is substantial - an interferometer with twice the arm length in current facilities, introducing novel technology only at the folding mirror. This upgrade path would be available for all existing terrestrial interferometers. Given additional facilities investment, this could be combined with an extension to interferometer arms, allowing commissioning and integration of folded interferometer systems while the original interferometer is in operation.

The singly-folded interferometer allows for an incremental introduction of cryogenics or novel coating materials to gravitational wave interferometers. Folded arm cavities only require cryogenic cooling for the folding mirror, which would be 4 km from the interferometer test masses, giving a potential reduction in noise coupling pathways. Incorporating cryogenic cooling into a gravitational wave interferometer requires significant engineering development to account for changes in materials at low temperatures and the noise impacts of cryocoolers.   The current manufacturing scale of AlGaAs coatings is sufficient for a folding mirror waist; the process of integrating smaller AlGaAs coatings into a folding mirror would provide the necessary research and development motivation for increasing the size of AlGaAs coatings in preparation for interferometers with larger beam sizes. 


\begin{figure*}
\includegraphics[width=2\columnwidth]{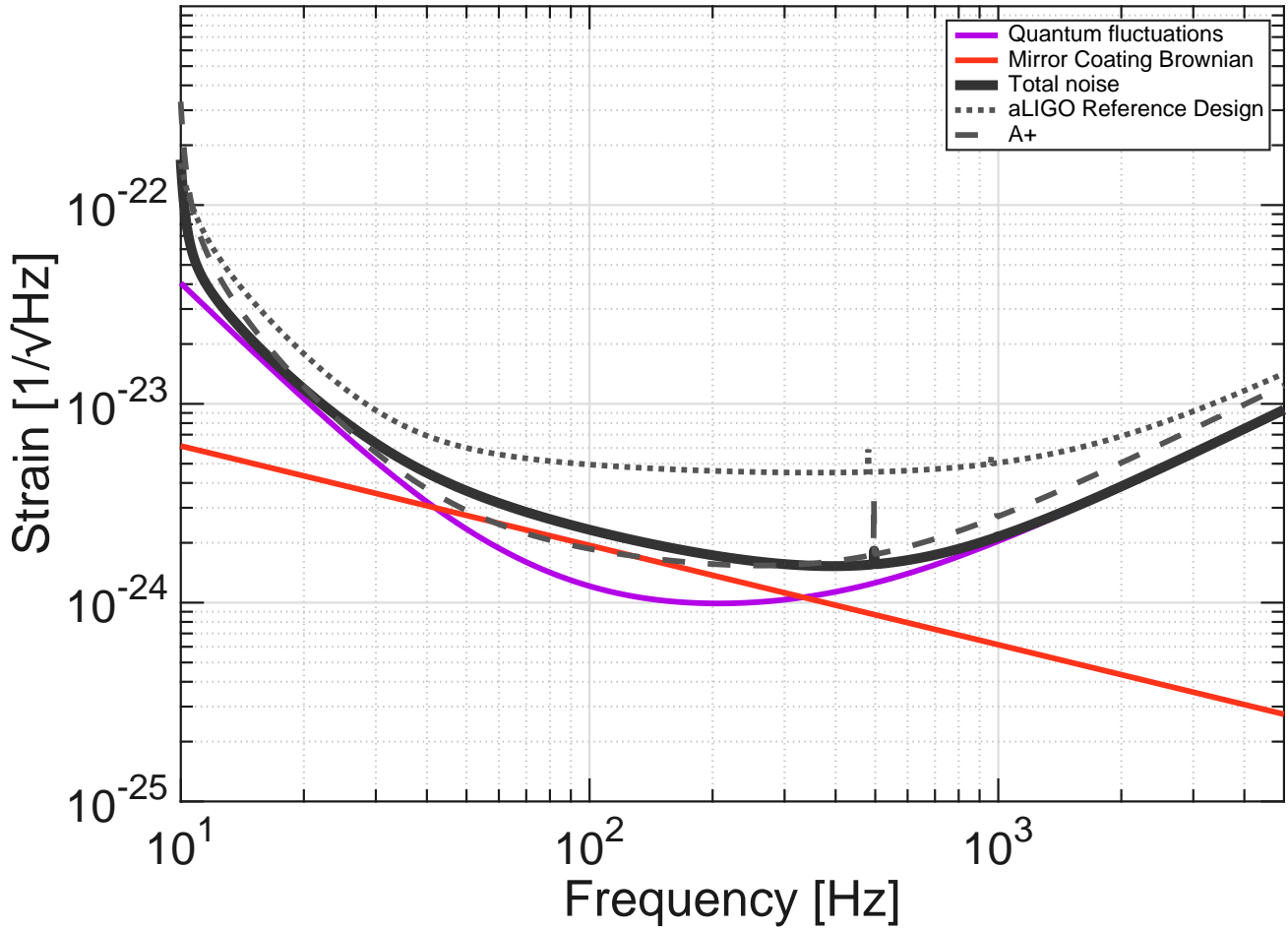}
\caption{Projected sensitivity curve for a singly-folded Fabry-Perot interferometer. The folding mirror has an AlGaAs crystalline coating, 9 dB of frequency-dependent squeezing is applied, and the signal recycling mirror transmittance is reduced from 0.2 to 0.16 to maintain a consistent interferometer bandwidth. All other technology is identical to the Advanced LIGO design. The Advanced LIGO design sensitivity is shown with a lighter dotted line. The A+ upgrade, assuming 9 dB of frequency-dependent squeezing and a reduction in coating thermal noise by a factor of two, is shown with a lighter dashed line.}
\label{fig:SensitivityProjection}
\end{figure*}

As the effective arm length increase from folding the arm cavities reduces the impact of many noise sources, as discussed in Dwyer et. al. \cite{PhysRevD.91.082001}, even a modest improvement to thermal noise can result in large range increases. Consider the simplest effective folded interferometer case, a folding mirror with an AlGaAs coating at room temperature, as a test configuration. The increased path length reduces the impact of quantum shot noise such that the interferometer noise curve is thermal noise dominated between 50 and 350 Hz. However, the folded configuration will double the impact of quantum noise due to radiation pressure, such that the interferometer will be limited by radiation pressure at low frequencies. Applying frequency-dependent quantum squeezing allows us to mitigate this radiation pressure increase without changing the size of the test masses. The range increase from a folded interferometer corresponds to a neutron star binary inspiral range of 375 Mpc and a black hole inspiral range of $z=2.2$, improving the aLIGO binary neutron star range by over 70\% and the aLIGO binary black hole range by over 85\%. The resulting noise curve is compared to the aLIGO design sensitivity and the A+ proposal for upgrading the Advanced LIGO instruments in Figure \ref{fig:SensitivityProjection}. Although A+ has similar range improvements and superior performance between 20 and 250 Hz, this instrument configuration requires a factor of two reduction in instrument thermal noise. \cite{InstrumentWhitePaper, Miller2015}. An investment in AlGaAs coating development is an investment in risk reduction, offering a potential increase in range on par with upgrades currently proposed for interferometers with unfolded interferometer arms. The sensitivity of folded interferometers can be further improved by increasing the mass of the test masses to reduce the impact of radiation pressure, cooling the folding mirror to further decrease the thermal noise, or applying any novel coating developed for unfolded interferometer cavities.


\section{Conclusion}
Folding existing Fabry-Perot arm cavities offers a simple upgrade path with significant range improvements for current interferometer facilities. Technology currently in development for major interferometer upgrades can mitigate the excess thermal noise from the folding mirror, allowing an increase in range of a factor of 2. This scheme allows for incremental introduction of new technology to the already successful advanced interferometer systems, reducing operational risk in the transition to future interferometers. The challenges in implementing this topology include improvements to controls systems for the three-mirror arm cavities and achieving the necessary thermal noise reduction with crystalline film coatings or cryogenics.

\section*{Acknowledgements}

The authors would like to thank Garrett Cole, Daniel Sigg, Lisa Barsotti, and Matthew Evans for fruitful discussions. This work was supported by the National Science Foundation grant PHY-1352511. This document has been assigned the LIGO Laboratory document number {LIGO-P1600231}.
   
\bibliography{fold}
\bibliographystyle{plain}
\end{document}